# MHITNet：a minimize network with a hierarchical context-attentional filter for segmenting medical ct images

Hongyang He[1*]

FENG Ziliang[2]

Yuanhang Zheng[2]

Shudong Huang[2]

haoBing Gao[2]

[1] College of Engineering and Physical Sciences, The University of Birmingham, Edgbaston Birmingham B15 2TT, United Kingdom；

\* Corresponding Author author's e-mail: HXH228@student.bham.ac.uk

[2]Institute of Image and Graphics, Sichuan Univ. Air Traffic Management College, Civil Aviation Flight Univ. of China

**Abstract**：In the field of medical CT image processing, convolutional neural networks (CNNs) have been the dominant technique.Encoder-decoder CNNs utilise locality for efficiency, but they cannot simulate distant pixel interactions properly.Recent research indicates that self-attention or transformer layers can be stacked to efficiently learn long-range dependencies.By constructing and processing picture patches as embeddings, transformers have been applied to computer vision applications. However, transformer-based architectures lack global semantic information interaction and require a large-scale training dataset, making it challenging to train with small data samples. In order to solve these challenges, we present a hierarchical contextattention transformer network (MHITNet) that combines the multi-scale, transformer, and hierarchical context extraction modules in skip-connections. The multi-scale module captures deeper CT semantic information, enabling transformers to encode feature maps of tokenized picture patches from various CNN stages as input attention sequences more effectively. The hierarchical context attention module augments global data and reweights pixels to capture semantic context.Extensive trials on three datasets show that the proposed MHITNet beats current best practises.

## 1 Introduction

Medical image segmentation is utilised to segment organs or lesions including retinal blood vessels [1–3], pelvic [4], polyp [5, 6], cardiac [7], cerebrum tumour [8–11], lung in CT pictures [12–14], nuclei in digital pathology images [15, 16], liver and tumour from CT volumes [17], and skin lesion [18]. Among all CNN variants used for medical image segmentation, the U-shaped networks demonstrate competitive performance. Lower-level semantic information is progressively downsampled through continuous convolutions and pooling layers to obtain high-level semantic information.However, it remains difficult for U-shaped approaches to increase their modelling efficiency in global situations.These end-to-end learning networks, including FCN [19], are comprehensive.and U-Net [20], utilise layered convolutions and sequential combining layers during encoding to obtain a suitably deep layers with expansive receptive fields. Even if skipconnections complement coarse-grained deep features with fine-grained features,local, fine-grained shallow detail information There will be a loss of functionalities and long-distance dependence.Edge feature is incapable of learning pixels and background.regions. The disadvantages of these omissions are themake the most of the present CNN-based performance

Deterioration of segmentation algorithms, such as inability to discover long-range relationships between pixels, insufficient attention to detail and erroneous boundaries. Alternatively,ResNet [21] is frequently seen as a backbone, however it is not.include the overall background and fine-grained specificswithout bypass connections Medical CT image segmentation can be broken down into three objectives: (1) determining existence; (2) rough segmentation of forms;boundaries.Recently, self-attention based architecture has becoming increasingly prevalent.

Transformers that are developed for sequence-to-sequence prediction have become the model of choice for natural language.processing (NLP) assignments [22] Transformer is a mechanism for self-focussed deep learning. Transformer's capacity to understand long-range relationships between input tokens is the key to their success. Consequently, transformer topologies often provide outstanding performance, particularly for target objects with a wide range of texture, shape, and size. Vision Transformer (ViT) [23] was introduced to accept 2D picture patches with positional embeddings as input sequence similar to NLP transformer approaches, and achieved competitive performance in computer vision applications. Data-efficient image Transformers (DeiT) were presented [24] to integrate knowledge distillation with image classification learning on a midsize dataset. By factorising 2D self-attention into two 1D self-attentions and a position-sensitive self-attention layer, Axial-Deeplab [25] produced stand-alone attention models with a broad or global receptive field. Segmentation Transformer (SETR) [26] was developed to merge transformer and CNN, in which transformer functioned as encoder and CNN acted as decoder, reaching a new state-of-the-art on ADE20K [27].We propose a hierarchical context-attention transformer network (MHITNet) for medical CT image segmentation, which is motivated by the analyses shown above. To investigate the linked information across scales, we construct a residual atrous pyramid pooling (RAPP) module in each skip-connection. The RAPP module sets different combinations of dilation rates according to the size of feature maps at various encoder stages, adapts the size of the feature map at the current stage, avoids overlapping of the functions of some branches, and obtains more diverse receptive fields and better multi-scale feature extraction capabilities than the ASPP [28] module with fixed dilation rate combinations of 1, 6, 12, and 18. In addition, residual connection is added to combine the original feature maps. Second, for utilising the transformer mechanism to acquire self-attention for multiscale samples, we employ a position-sensitive axial attention (PAA) module following the RAPP module at each stage of skip-connection. Thirdly, we build a hierarchical context-attention (HCA) module to compensate for the global context-attention information lost as a result of the long-range linkages between visual patches depicted by transformer mechanism.

Three modules are cascaded in skip-connection to discover the multi-scale, long-range interdependence and contextual attention of medical CT image segmentation targets. Compared to pure CNNs, it augments the modelling of long-range dependencies between pixels; compared to transformer-based networks, it utilises the transformer mechanism as self-attention, without affecting the backbone network's ability to capture local details, thereby mitigating the disadvantages of the transformer mechanism in capturing local details.

## 2.Methodology

Figure 1 depicts the structure of the MHITNet. It consists largely of three modules: the RAPP module, the PAA module, and the HCA module. We use ResNet-34 [21] as the foundation of our encoder and bilinear interpolation as our decoder. The RAPP module, the PAA module, and the HCA module are added progressively to each skip-connection.For each skip connection, we employ a summing operation as opposed to a concatenation operation in order to raise the difference between the anticipated foreground and background values and to improve the performance capabilities of the suggested modules. The proposed MHITNet uses pre-trained ResNet-34 [21] with a large amount of data as its backbone, retaining the local feature capture ability of CNN and fusing feature maps on skip-connections with feature maps of the backbone to avoid the poor training effect of transformer-based architecture on the dataset of medical images with limited data samples. In addition, we employ summation operation.

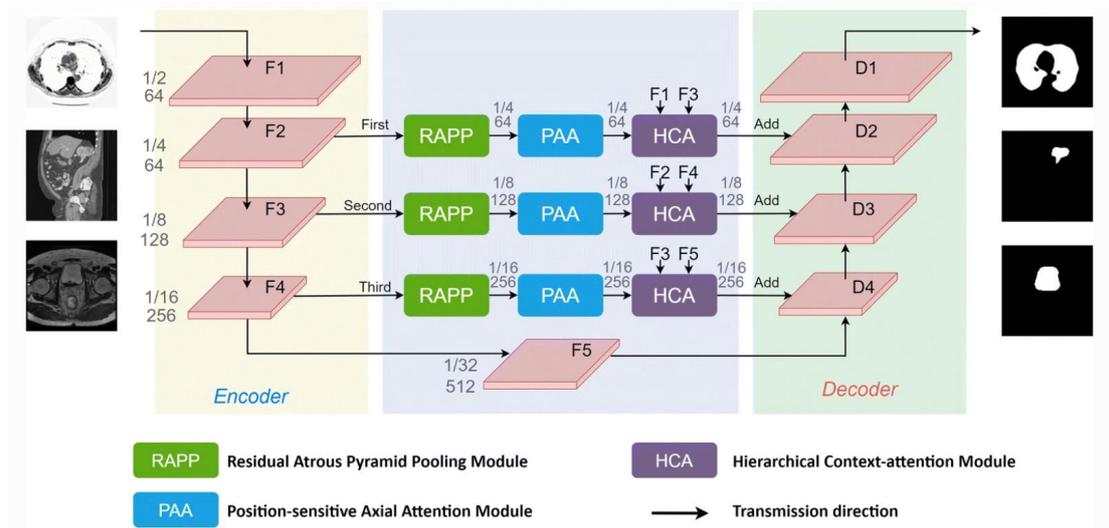

Figure 1.The framework of the MHIT-Net model

### 2.1 The RAPP module

Different-sized lesions give valuable information regarding the shape and size of target items. Encoder in U-Net [20] naturally includes layers of continuous convolutions and max-pooling. Consequently, visual resolution diminishes continually during downsampling, and information becomes increasingly abstract. Due to the detail loss induced by double convolutions and downsampling, we construct a RAPP module to investigate the multi-scale correlated information. As seen in Fig. 2, the plug-and-play RAPP module contains five parallel convolution branches and one preceding feature branch that provide the same number of channels and size of feature maps. The first branch uses an adaptive average pooling layer to convert the input H W Cout feature map to a 1 1 Cout feature map, the concatenated 11 convolutional layer performs a channel fusion operation to obtain the 1 1 Cout feature map, and then bilinear interpolation is used to restore the original H W Cout feature map. When the sampling rate is near to the size of the feature map, the first branch's purpose is to gather image-level features to compensate for filter deterioration. The second through fifth branches are each composed of atrous convolutions. In order to combine the acquired multi-scale features and preserve more information, we concatenate them and apply an 11 convolution to minimise their dimension.has five parallel convolutional branches and one preceding feature branch that generates same sized feature mapsas well as channel counts.The four branches of the ASPP [28] establish fixed atrous rate combinations of 1, 6, 12, and 18 or 6, 12, 18, and 24. Nevertheless, the receptive field formed by these combinations is significantly bigger than the objective when learning small size feature maps, such as 1616, resulting in a reduction in the capacity of branches to acquire varied features and overlapping effects. In contrast to the ASPP module, which uses the same atrous rate combination across all stages, the RAPP module is developed with various dilation rates based on input feature maps of varying sizes in skip-connection to adaptively learn numerous scales more efficiently. In the initial skip-connection, the dilation values are set to 1, 6, 12, and 18 from top to bottom. Setting the second dilation values to 1, 3, 5, 7 and the third dilation values to 0, 1, 2, 4.

**2.2 The PAA module**

The PAA module is designed to capture long-range dependencies among the pixels of an image. Following [23], a trainable neural network can be built by stacking transformers. The transformer module is composed of self-attention and feed forward neural network.

Self-attention mechanism calculates similarity by the same sample and then predicts the weight of other parts. Specifically, the input $X \in R^{H \times W \times C_{in}}, X \in R^{H \times W \times C_{in}}$ of feature map, the output $Y \in R^{H \times W \times C_{out}} Y \in R^{H \times W \times C_{out}}$ of self-attention is computed as follows:

$$Y = \text{Softmax}\left(Q \times K^T / \sqrt{d_k}\right) \times V \tag{1}$$
$$Q = W_q \times X, K = W_k \times X, V = W_v \times X \tag{2}$$

axial attention [25, 42] divides self-attention into height-axis multi-head attention and width-axis multi-head attention modules to solve the shortcomings of self-attention mechanism. The height-axis multi-head attention does self-attention between each pixel and all pixels in the same height-axis, while the width-axis module performs self-attention between each pixel and all pixels in the same width-axis. These two modules enhance the position deviation to make the similarity calculation responsive to location data. The formula for the modified position-sensitive attention along the width is

(3): 18, respectively, from top to bottom. Setting the second dilation values to 1, 3, 5, 7 and the third dilation values to 0, 1, 2, 4.

$$y_{ij} = \sum_{w=1}^{W} \text{softmax}\left(q_{ij}^T k_{iw} + q_{ij}^T r_{iw}^q + k_{iw}^T r_{iw}^k\right)(v_{iw} + r_{iw}^v) \tag{3}$$

Note that Eq. 3 applies position-sensitive axial consideration along the width axis. Height-axis axial attention is likewise accounted for using a similar technique. Since the transformer mechanism has the benefit of great computing efficiency for accurate distant interactions [25], the combination of height-axis and width-axis axial attention constitutes a suitable model of self-attention.

Combining convolutions with axial self-attention transformers, our module generates position-sensitive axial attention. As illustrated in Figure 3, the input X of the PAA module is the output of the RAPP module for multi-scale learning. Initially, input X yields an output feature map x that captures local features by double convolutions. Second, x is sequentially subjected to an 11 convolution, a height-axis multi-head attention module, a width-axis multi-head attention module, and an 11 convolution, with the previous feature x added through residual connection to generate the attention map. Thirdly, attention map is smoothed by an 11 convolution and preceding feature x is recombined by concatenation to bring self-attention closer to the ground-truth for improved medical CT image segmentation. In addition, a bilinear upsampling process yields the final attention map.

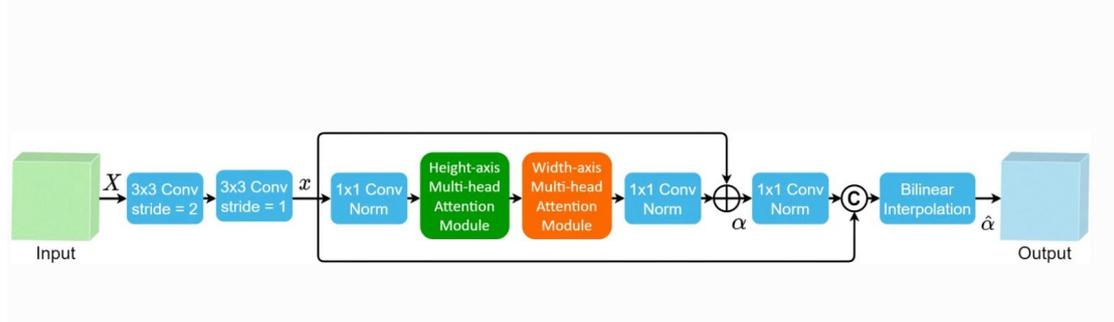

Figure 2.The illustration of the PAA module

## 3.Results and Discussion
### 3.1Experiment Environment and Metrics

Precision and Recall are used to objectively evaluate the experimental data by measuring the performance of the model using an IOU threshold of 0.5 and a confidence threshold of 0.4. Precision is the chance of accurately predicting a positive sample from among all projected positive samples, whereas recall is the probability of correctly predicting a positive sample from among the actual positive samples. Formulae (4), (5), TN (predict negative samples as negative samples), FN (predict positive samples as negative samples), TP (predict positive samples as positive samples), and FP display the precision and recall formulas (predict negative samples as positive samples).

$$P = \frac{TP}{TP + FP} \tag{4}$$

$$R = \frac{TP}{TP + FN} \tag{5}$$

Precision and Recall interact in object detection and cannot be used alone to assess the detection. Therefore, we offer the AP to describe the detection precision and the F1-Score to assess the model in a more complete manner. Higher AP and F1-Score values indicate a more accurate network, whereas the mAP shows the mean accuracy for n kinds of faults. The AP, mAP, and F1-Score formulae are illustrated in formulas (6), (7).

$$AP = \int_0^1 P(R)dR \tag{6}$$

$$mAP = \frac{1}{n}\sum_{i=1}^{m} AP^i \tag{7}$$

$$F1 - Score = 2 \times \frac{P \times R}{P + R} \tag{11}$$

When evaluating the superiority of the algorithm we need to define the detection speed, here we use the frame rate (FPS) to indicate the detection speed, which is an important indicator, if the FPS ≥ 30, it satisfies the requirements, and a video detection function with FPS ≥ 60 is superior.

**3.2 Experimental settings**

The suggested MHITNet is based on the PyTorch framework and is trained using one NVIDIA RTX 3060 graphics card and 12 gigabytes of RAM.

Our model utilised the Adam optimizer for 300 iterations. 1 and 2 are assigned the values 0.9 and 0.999, respectively. Experiments with equal resolution employ the same number of samples each batch. During training, the initial learning rate was set at 10 4 and degraded by a factor of 0.75 every 20 epochs.

Lung CT image segmentation

MHITNet is compared to numerous cutting-edge algorithms, such as U-Net [20], AG-Net [30], CE-Net [37], Axial Attention U-Net [25] and MedT [42]. Note that Axial Attention U-Net and MedT employ the binary cross-entropy loss function.

MHITNet provides the best AC performance with a resolution of 128128 pixels, which is increased by 0.13%, 0.24%, 0.3%, 4.41 %, and 4.56% compared to U-Net, AG-Net, CE-Net, Axial Attention U-Net, and MedT on the lung dataset. In terms of SE, our model delivers 10,93% and 11,21% advantages over the most recent transformer-based medical picture segmentation techniques, Axial Attention U-Net and MedT, respectively. The transformer-based technique replaces a portion of the convolution layer used for feature extraction by incorporating a transformer mechanism into the encoder. Its capacity to segment a pixel-level, single CT target is restricted because to its inability to collect local characteristics.

| Dataset | Lung | | | | KiTS19 | | | |
| --- | --- | --- | --- | --- | --- | --- | --- | --- |
| Methods | AC | SE | AUC | DS | AC | SE | AUC | DS |
| U-Net [20] | 0.9881 ± 0.0269 | 0.9516 | 0.9607 | 0.9402 | 0.9899 ± 0.0254 | 0.7882 | 0.8937 | 0.9183 |
| AG-Net [30] | 0.9887 ± 0.0264 | 0.9625 | 0.9796 | 0.9689 | 0.9911 ± 0.0248 | 0.8501 | 0.9240 | 0.9219 |
| CE-Net [37] | 0.9898 ± 0.0255 | **0.9811** | 0.9867 | 0.9753 | 0.9915 ± 0.0245 | 0.8406 | 0.9197 | 0.9331 |
| Axial Attn U-Net [25] | 0.9470 ± 0.0553 | 0.8715 | 0.9221 | - | 0.9706 ± 0.0393 | 0.7237 | 0.8106 | - |
| MedT [42] | 0.9455 ± 0.0593 | 0.8687 | 0.9191 | - | 0.9745 ± 0.0372 | 0.7575 | 0.8772 | - |

Table 1 Segmentation comparisons on lung and KiTS19 datasets with a resolution of 128 × 128

In addition, MHITNet delivers state-of-the-art performance in most measures with 512512 pixel resolution. Among all the data shown in Table 1, the AC of MHITNet is 99.48%, which is an increase of 1.98 percentage points and 0.46 percentage points, respectively, compared to U-Net and CE-Net. It has been noticed that the design of fusion transformers and convolution-based modules introduced in skip-connection precedes that of pure CNNs or transformer-based encoder algorithms. features.

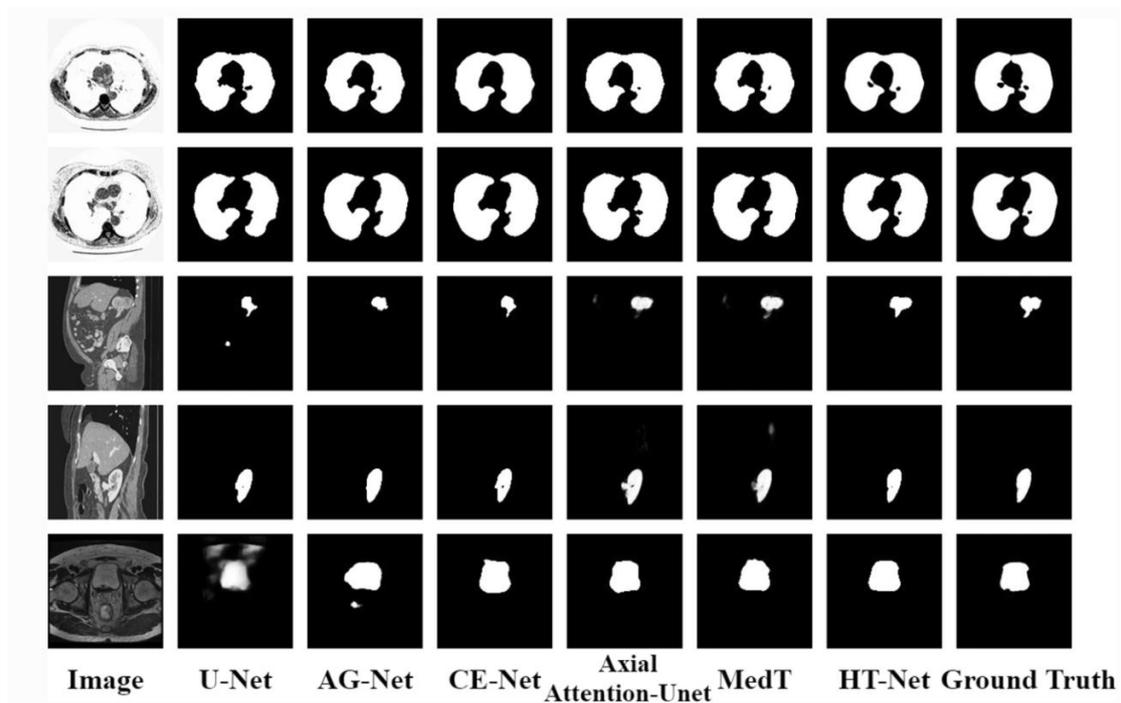

Figure3.Visual segmentation comparisons of MHITNet with five state-of-the-art methods on the lung, KiTS19 and bladder CT datasets

**3.3CT segmentation of the KiTS19 Kidney Tumor**
The MHITNet surpasses all five cutting-edge techniques with a resolution of 128128, as seen in Table 1. MHITNet obtains AC, SE, AUC, and DS values of 99.24%, 88.49%, 94.13 %, and 93.75%, respectively, on the KiTS19 dataset, as demonstrated by the experimental results. In contrast to the findings of traditional U-Net and the top rival CE-Net, the AC increases by 0.25 percent and the SE grows by 9.67 percent and 4.43 percent, respectively. Regarding the evaluation transformer-based baselines of AC with the most recent Axial Attention U-Net [25] and MedT [42], MHITNet is enhanced by 2.18 and 1.78 percent, respectively. MHITNet consistently beats all SOTAs on Table 2 at 512512 resolution.

In terms of SE and AUC assessment measures, Axial Attention U-Net and MedT achieve exceptionally poor results. During the segmentation of microscopic targets, we hypothesise that the similarity of 2D CT image patches is computed using a transformer-based baseline, which is influenced by a great deal of background noise.

In the third and fourth rows of Figure 3, the visual segmentation results of our technique outperform those of other methods.

Bladder CT image segmentation

Table 2 displays comparative findings for the bladder dataset. The MHITNet surpasses the best convolution-based competition CE-Net and the most recent transformer-based baseline MedT by 0.11 percent and 2.25 percent, respectively, in terms of the AC. And MHITNet again obtains the greatest SE performance, with gains of 1.79% and 25.88% above CE-Net and MedT, respectively.

| Dataset | Lung | | KiTS19 | |
| --- | --- | --- | --- | --- |
| Methods | AC | DS | AC | DS |
| U-Net [20] | 0.9750 ± 0.0322 | 0.9592 | 0.9902 ± 0.0240 | 0.9473 |
| Backbone | 0.9909 ± 0.0231 | 0.9754 | 0.9920 ± 0.0222 | 0.9586 |
| Backbone+RAPP | 0.9927 ± 0.0197 | 0.9783 | 0.9923 ± 0.0219 | 0.9596 |
| Backbone+PAA | 0.9932 ± 0.0193 | 0.9828 | 0.9930 ± 0.0213 | 0.9599 |
| Backbone+HCA | 0.9935 ± 0.0191 | 0.9860 | 0.9933 ± 0.0211 | 0.9604 |
| Backbone+RAPP+PAA | 0.9942 ± 0.0187 | 0.9865 | 0.9939 ± 0.0208 | 0.9611 |
| Backbone+PAA+HCA | 0.9945 ± 0.0185 | 0.9867 | 0.9936 ± 0.0210 | 0.9614 |
| Backbone+RAPP+HCA | 0.9944 ± 0.0184 | 0.9871 | 0.9940 ± 0.0206 | 0.9605 |

Table 2.Ablation analysis for the MHITNet on lung and KiTS19 datasets with a resolution of 512 × 512

With the inclusion of the RAPP module to the backbone network, edge details of objects are discriminated in visual attention maps, and the network gets the capacity to concentrate on learning multi-scale object and edge detail information.

The PAA module: With the inclusion of a distinct PAA module, the network offers a basic sketch of the organ and learns the region of interest via self-attention and long-range interdependence. We see, however, that the PAA module does not learn well in the lung dataset's abstract feature map. Consequently, we run further tests on various skip-connections.

Figure 4 depicts a visual comparison of the PAA module on three skip-connections. In the lung (512512) and KiTS19 (128128) datasets, the shallow abstract feature map has a more precise interest region than the deep abstract feature map. We believe that adequate pixel computation in the same dataset yields favourable axial similarity information, and that the learning performance of axial self-attention is enhanced in high resolution feature maps.

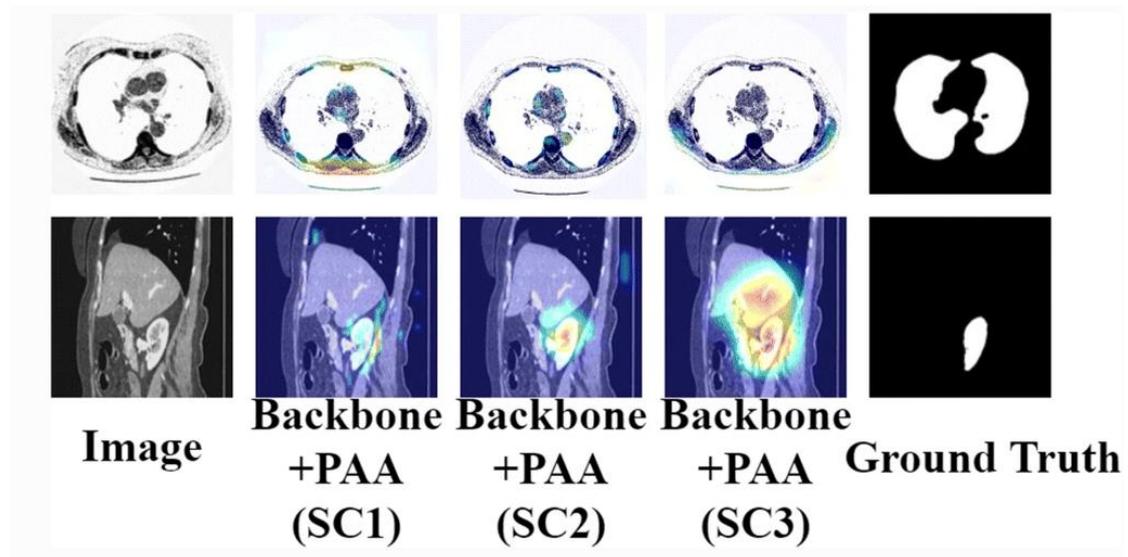

Figure4.Examples of visual attentional map comparison of the PAA module on three skip-connections. The SC1, SC2, and SC3 represent the first, second, and third skip-connection, respectively

Combinations of the suggested modules: Various combinations of the proposed modules can provide distinct results. For instance, a combination of the suggested RAPP and PAA modules focuses the network's attention on multi-scale organ learning and diverts it from other objects. A combination of the proposed PAA and HCA modules is superior to a combination of the proposed RAPP and HCA modules at focusing on a specific subset of organs. In conclusion, MHITNet, which is comprised of the suggested RAPP, PAA, and HCA modules, provides more precise segmentation focus, and its zone of interest is highly congruent with the ground truth.

## 4.conclusion
Combining convolutions and transformers, MHITNet accurately segments lesions in medical CT images. In our study, we suggest three cascading components. Using the suggested RAPP and PAA modules, MHITNet is able to acquire boundary information at many scales and simulate long-distance pixel relationships. In addition, the HCA module is implemented to extract context information from hierarchical encoding stages and reweight pixels based on context-attention. The suggested convolution modules can effectively compensate for the shortcomings of the transformer technique for collecting local characteristics. Numerous experimental findings demonstrate that MHITNet outperforms the state-of-the-art on three public CT imaging datasets using four frequently used performance measures. We will continue to refine the model and investigate the hybrid network of convolutions and transformers in the future.

# 5.reference